# Contribution de Hervé Suaudeau lors de son audition par la mission d'information de la commission des lois du Sénat consacrée au vote électronique le 25 janvier 2018

Hervé Suaudeau
Membre expert de l'Observatoire du Vote
herve.suaudeau@parisdescartes.fr

**Résumé :**

**Cette contribution étudie la compatibilité entre l'usage de scrutins dématérialisés et l'article 3 de la constitution qui dispose que « le vote doit être toujours universel, égal et secret ». La gestion du risque informatique du vote électronique se mène de façon rationnelle et pragmatique comme tout autre domaine de gestion de risque. Nous sommes donc prêts à accepter d'utiliser une technique non dématérialisée pour couvrir un risque important, bien identifié et non évitable. Or, les recherches ont démontré qu'un un vote dématérialisé vérifiable perd intrinsèquement son anonymat. De plus l'étude de l'historique des failles de sécurité démontre qu'aucun terminal de ces dernières années n'a pu garantir l'anonymat de sa connexion et que les données personnelles de vote ont toujours été potentiellement exposées. Les systèmes de vote électroniques sont donc, dans l'état des connaissances actuelles, incapables d'apporter la garantie constitutionnelle requise.**

**Abstract :**

**This contribution investigate the compatibility between the use of dematerialised voting and Article 3 of the french Constitution, which stipulates that "voting must always be universal, equal and secret". The IT risk management of electronic voting is driven in a rational and pragmatic way like any other risk management area. We are therefore prepared to accept the use of a non dematerialised technique to cover a significant, well-identified and unavoidable risk. However, research has shown that a verifiable dematerialized vote intrinsically loses its anonymity. In addition, a study of the history of vulnerabilities shows that no terminal in recent years has been able to guarantee the anonymity of its connection and that personal voting data has always been potentially exposed. Electronic voting systems are therefore, to the best of our knowledge, unable to provide the required constitutional guarantee.**

*NB : Les titres des chapitres, le chapitre « COMPLEMENTS », les notes et références de bas de page n'ont pas été lus in extenso lors de cette audition mais plutôt évoqués au travers des différentes questions des parlementaires.*

Je suis Hervé Suaudeau, ingénieur en informatique et en électronique et j'interviens ici – en remplacement de M. Graton empêché – en qualité de représentant de l'Observatoire du Vote, institut indépendant dont la vocation est l'éducation au vote, l'observation des élections et l'analyse de leurs dispositifs.

Je suis également administrateur système et réseau informatique au Centre National de la Recherche Scientifique.

**La garantie constitutionnelle du suffrage et du vote en France**

Commençons directement au but de ma contribution :

La constitution dispose dans son article 3 que (je cite) « Le suffrage […] est **toujours** universel, égal et **secret** ». Ici, nous conviendrons facilement que l'utilisation du terme « toujours » impose au législateur de mettre en place un cadre qui ne souffre d'aucune ambiguïté.

La constitution impose ainsi à ce qu'il soit garanti par la loi et l'État que chaque électeur ait droit à ce que son suffrage personnel soit effectivement pris en compte et que son vote soit bien secret.

Or qu'en est-il en ce qui concerne le vote dématérialisé par l'informatique et comment approcher cet objet pour des personnes comme vous qui ne peuvent être spécialistes de tout, notamment concernant les connaissances technologiques et scientifiques qui font consensus dans le monde de la sécurité informatique auquel s'applique le vote par internet ?

La présente contribution va donc aborder ces questions de manière la plus exacte possible tout en restant pédagogique et se maintiendra sur le terrain des connaissances, prenant la précaution de demeurer éloigné des idéologies, et des discours des industriels ou des prestataires de services.

**Le métier de la sécurité informatique au quotidien**

Pour commencer, je ne souhaite pas dresser un tableau alarmiste de la sécurité informatique, car c'est un domaine qui se gère et doit se gérer, comme d'autres, de façon sereine, rationnelle et pragmatique. Il s'agit ici de surtout pointer les spécificités de ces questions pratiquées au quotidien par les directions de sécurité informatiques du pays. En aucune façon un de leurs responsables sérieux ne pourra vous garantir la disparition des risques informatiques. En effet, leur métier, que j'ai la chance de pratiquer, est de faire baisser ce risque à un niveau acceptable. Cette baisse est opérée au regard 1) des efforts consentis pour éviter le risque, 2) des probabilités d'occurrence des incidents et surtout 3) de leurs conséquences potentielles. Parfois, les seules mesures efficaces de défense face à ces risques consistent à débrancher du réseau un système ou un service n'ayant pas de caractère impératif à y être connecté. D'autres fois, même, il est demandé d'utiliser une autre technique que l'informatique pour accomplir la tâche en question.

**L'impossible anonymat théorique du vote dématérialisé**

Il faut donc avoir conscience que la remédiation d'un risque informatique ne passe pas toujours par une solution informatique.

En effet, il existe certaines tâches, qui ne peuvent être dématérialisées sans ce qu'on pourrait appeler des « effets secondaires » inévitables. Bien que certains professionnels ne le sachent pas, le vote anonyme, est démontré faire partie de cette catégorie depuis 2006, quand une équipe de chercheurs français – excusez la longue citation, mais cette publication publiée en 2006 et 2010[1] est majeure et je tiens à les citer – Benoît Chevallier-Mames de Gemalto, Pierre-Alain Fouque et David Pointcheval de l'ENS et du CNRS, Julien Stern de Crytolog International et Jacques Traoré de France-Télécom R&d (nom à l'époque) a démontré mathématiquement la chose suivante : un vote dématérialisé, ne peut être à la fois anonyme et vérifiable par l'électeur[2]. Il n'y a ainsi pas

---
[1] Benoît Chevallier-Mames, Pierre-Alain Fouque, David Pointcheval, Julien Stern et Jacques Traoré. On some incompatible properties of voting schemes. Towards Trustworthy Elections, 6000 :191–199, 2010.
[2] La formulation exacte de l'article scientifique dans reformulé dans la reprise de leur article en 2010 est la suivante (traduction personnelle) « En conclusion, nous avons montré que les systèmes de vote ayant les caractéristiques habituelles ne peuvent pas vérifier les notions de sécurité fortes toutes en même temps : nous ne pouvons pas

d'échappatoire : un vote dématérialisé anonyme, ne peut être garanti vérifiable et inversement, un vote dématérialisé vérifiable, ne peut être garanti anonyme. Pour parler plus concrètement personne ne peut vérifier ce qui a été voté informatiquement, et inversement si un vote dématérialisé est vérifiable, c'est qu'il n'est pas anonyme.

Cette démonstration scientifique pose fondamentalement la question de la constitutionnalité du vote par internet.

**Le vote dématérialisé est affecté comme tout autre processus par les failles informatiques**

Mais au-delà de cette démonstration scientifique, intéressons-nous également concrètement aux conditions dans lesquelles se pratique le vote par internet :

Le vote par internet est opéré sur des terminaux et des serveurs, autrement dit, le vote est effectué sur des ordinateurs ou des smartphones connectés d'un bout à l'autre par internet. Or ceux-ci sont soumis à de nombreuses failles informatiques qui peuvent mettre en danger l'anonymat ou la sincérité de ce vote.

En effet, chaque année des milliers de cybermenaces sont révélées et il faut que ce groupe de travail prenne conscience de leur ampleur et de leur progression. Ainsi en 2017, en augmentation de +70 % par rapport à l'année précédente, 18 511 vulnérabilités ou soupçon de vulnérabilités informatiques dans les outils que nous utilisons tous les jours ont été publiées et officiellement identifiées par les autorités (dans le catalogue de l'organisme MITR chargé de les archiver)[3]. Toutes ne donnent pas lieu heureusement à des alertes graves, mais ce matin encore, le site du CERT-FR (Centre gouvernemental de veille, d'alerte et de réponse aux attaques informatiques) recense 4 alertes majeures en cours[4] pour lesquels aucune solution définitive ou satisfaisante n'a été correctement publiée.

Ces alertes ont une conséquence directe sur le vote électronique et nous devons prendre un exemple pour l'illustrer : Vous avez peut-être entendu parler dans les grands médias ces jours-ci de SPECTRE, une vulnérabilité particulièrement inquiétante. Elle a été rendue publique le 3 janvier dernier et, après une hésitation pendant quelques jours, on sait maintenant qu'elle concerne pratiquement tous les appareils connectés[5]. Elle se loge au cœur du fonctionnement des cerveaux de nos ordinateurs et smartphones. C'est une faille au niveau de la conception des processeurs qui ne peut pas être comblée par une mise à jour de leur logiciel. Cette faille permet à un utilisateur malveillant de consulter en clair, toutes les données qui sont utilisées au sein de l'ordinateur, données auxquelles il ne devrait pas avoir accès. C'est très problématique, car un programmeur malveillant peut ainsi récupérer les mots de passe, les codes de chiffrement et les données qui sont censées rester secrètes dans un ordinateur comme les votes ou les identifiants de vote si celui-ci est utilisé pour un suffrage électronique. Au passage, il est intéressant de savoir pourquoi cette faille porte le nom de « spectre » : Les ingénieurs de Google qui ont révélé la vulnérabilité ont tout de

---

parvenir simultanément à la vérifiabilité universelle du décompte des voix et à la confidentialité inconditionnelle des votes ou à la 'receipt-freeness' » (*l'absence de preuve de son vote protège l'électeur contre la coercition*). Texte original : « As a conclusion, we have shown that voting systems with usual features cannot simultaneously achieve strong security notions : we cannot achieve simultaneously universal verifiability of the tally and unconditional privacy of the votes or receipt-freeness. »

3   Toutes les descriptions des vulnérabilités sont téléchargeables année par année sur le catalogue du MITRE
    http://cve.mitre.org/data/downloads/
4   Les alertes en cours à la date du 25/01/2018 sont CERTFR-2017-ALE-018, CERTFR-2017-ALE-019, CERTFR-2017-ALE-020 et CERTFR-2018-ALE-001 https://www.cert.ssi.gouv.fr/
5   https://www.cert.ssi.gouv.fr/alerte/CERTFR-2018-ALE-001/

suite compris qu'elle serait très compliquée à contrer et qu'ainsi elle viendrait nous hanter pendant longtemps, telle un spectre[6].

Le plus grave dans cette faille n'est pas qu'il soit très difficile de s'en prémunir. Le plus grave est que celle-ci existe depuis au moins 15 ans[7] et qu'il est impossible d'affirmer qu'elle n'a pas été utilisée durant cette période.

En effet, il est courant de découvrir des vulnérabilités qui sont utilisées, parfois massivement et durant de longues périodes, avant qu'elles ne soient révélées. Parmi les plus connues, la dénommée HeartBleed a été utilisé massivement par des pirates au moins 5 mois avant sa révélation[8]. Cette faille permettait, durant une période théorique de deux ans, à des internautes mal intentionnés de récupérer à l'insu de l'utilisateur des informations personnelles qui circulaient de manière chiffrée sur internet. Les protocoles de vote utilisant ce chiffrement, réputé sûr jusqu'en avril 2014, étaient en fait totalement exposés et pouvaient révéler de manière indétectable leurs contenus. On estime par exemple qu'à l'époque, en plus des nombreux serveurs, des routeurs et équipements d'opérateurs et de centres serveurs, que 50 millions de smartphones Android étaient touchés[9]. Bien que à partir de cette date, nombre de précautions aient été prises dans la conception de ces logiciels, des nouvelles failles similaires en gravité portant sur les mêmes logiciels utilisés pour sécuriser le trafic internet ont quand même été découvertes fin 2014 et début 2016 (failles aux noms de « Poodle » et « Drown »). Aux vues de ces révélations, personne ne peut garantir la fiabilité ou l'anonymat d'une communication internet et il n'est pas d'avantage possible de garantir la détection de l'exploitation de ces failles lorsqu'elles sont mises en œuvres.

Les outils pour casser l'anonymat de nos connexions internet circulent parfois pendant longtemps avant d'être révélés au grand public. Le cas « EternalBlue » révélé le 14 avril 2017, est un outil qui a été utilisé par l'agence de sécurité américaine NSA durant 5 ans avant que leur logiciel ne fuite[10] et soit repris par des pirates pour être intégré dans des virus tels que WannaCry et NotPetya entraînant à eux deux des centaines de millions d'Euros de dommages, tel que la presse l'a rapporté.

Plus grave, nombre de terminaux, particulièrement parmi les smartphones, ne sont jamais mis à jour pour corriger les failles de sécurité[11]. Alors que la majorité des Français utilisent désormais leur mobile pour se connecter à Internet[12], les fabricants et opérateurs de nombreux téléphones rechignent à produire des mises à jour et laissent leurs utilisateurs à la merci de failles dont on sait pourtant parfaitement se prémunir. Ainsi des millions de citoyens de notre pays utilisent au quotidien des terminaux dont on a déjà la preuve et la démonstration qu'ils ne peuvent aucunement garantir l'anonymat de leur connexion.

---

6   « As it is not easy to fix, it will haunt us for quite some time. » https://spectreattack.com/
7   La faille Meltdown, révélée en même temps que Spectre, avec les mêmes effets, mais plus facilement corrigeable, est présente dans la plupart des processeurs depuis 1995.
8   Sean Gallagher, « Heartbleed vulnerability may have been exploited months before patch [Updated] », Ars Technica, 9 avril 2014 https://arstechnica.com/information-technology/2014/04/heartbleed-vulnerability-may-have-been-exploited-months-before-patch/
    Peter Eckersley, « Wild at Heart: Were Intelligence Agencies Using Heartbleed in November 2013? », Electronic Frontier Foundation, 10 avril 2014 https://www.eff.org/deeplinks/2014/04/wild-heart-were-intelligence-agencies-using-heartbleed-november-2013
9   Charles Arthur « Heartbleed makes 50m Android phones vulnerable, data shows » The Guardian, 15 avril 2014, https://www.theguardian.com/technology/2014/apr/15/heartbleed-android-phones-vulnerable-data-shows
10  « 'NSA malware' released by Shadow Brokers hacker group », BBC News, 10 avril 2017, http://www.bbc.com/news/technology-39553241
11  Julien Cadot  « Sécurité sur Android : les mises à jour critiques sont toujours massivement ignorées », Numérama, 23 mars 2017, https://www.numerama.com/tech/242932-securite-sur-android-les-mises-a-jour-critiques-sont-toujours-massivement-ignorees.html
12  Nicolas Richaud, « Les Français utilisent en majorité leurs smartphones pour se connecter à Internet », Les Echos, le 28/09/2017, https://www.lesechos.fr/28/09/2017/lesechos.fr/030631696692_les-francais-utilisent-en-majorite-leurs-smartphones-pour-se-connecter-a-internet.htm

Toutes ces failles et ces nouvelles, qui aujourd'hui font les grands titres de presse, ne surprennent pas les professionnels de la sécurité informatique qui ont maintenant une trentaine d'année de recul sur la question. Anecdote intéressante, les CERT, centre d'alerte et de réaction aux attaques informatiques dans le monde, ont été créés suite à l'infection en novembre 1988 – infection ici accidentelle – de 3 à 4 % des machines connectées à internet à l'époque, rendant inutilisable une bonne part du réseau[13]. Et depuis, découvrir l'existence de faille informatique est devenu notre quotidien.

**Conclusion**

En conclusion, l'informatique, et ce n'est pas une grande révélation, démontre sa faillibilité de manière permanente. Les failles qui sont identifiées de nombreuses années après leur apparition, et même de nombreuses années après leur utilisation, apportent la démonstration que l'ensemble des systèmes n'est pas garanti.

Le vote électronique est un domaine radicalement différent des autres domaines de l'informatique par exemple le domaine bancaire, où chaque opération peut être vérifiée atomiquement du fait de l'absence d'anonymat. En vote électronique, qu'il soit par internet où sur machine à voter, ce n'est pas le cas. Ainsi en vote électronique, il y a un problème à chaque fois qu'est nommé un expert qui ne pointe aucune faille de sécurité (et comment pourrait-il pointer des failles dont il n'a pas encore connaissance et dont la découverte – toujours à posteriori – nécessite des dizaines d'experts de très haut niveau dans le monde ?). L'expert ne peut d'aucune manière être une garantie. En France aucun d'entre eux chargé de vérifier la conformité d'un système de vote aux contraintes légales et constitutionnelles n'a jamais détecté les failles révélées par la communauté quelques mois, voire quelques années plus tard et qui pourtant rendaient le dispositif audité impropre à la garantie des droits requis. Pour ne prendre qu'un seul exemple les élections législatives de 2012 des représentants des Français établis hors de France qui se sont déroulées en partie par Internet, étaient exposée (et pour ne prendre que les exemples que j'ai cité) aux failles Drown, Poodle, EternalBlue et bien sûr Spectre. Je passe nombre d'élections dont les centaines de scrutins annuels professionnels qui se déroulent par internet chaque année. Il est ainsi aujourd'hui démontré, que pour toutes ces élections, où pourtant un expert a été nommé, l'anonymat a pu être violé y compris sur les ordinateurs sains et parfaitement mis à jour à l'époque. J'espère que les personnes de cette mission ont bien compris le sens de cette phrase.

**Conclusion**

L'étude de législation comparée du Sénat en 2007 sur cette question alertait déjà, alors que les connaissances révélées sur les failles de sécurité étaient bien moindre. Sécuriser l'anonymat d'un vote n'est pas une question de croyance, c'est une garantie impérative constitutionnelle à laquelle on ne peut pas se soustraire.

Si l'on se place au niveau de l'organisation des élections en amont, nous n'avons ainsi pas le droit d'organiser des scrutins avec des bulletins dont on sait que sous certaines conditions ils peuvent faire apparaître leur contenu, être modifiés ou ne jamais être pris en comte. Concernant le vote papier, je pourrais revenir sur des détails si vous le souhaitez, mais la continuité matérielle du procédé et la destruction des bulletins après le scrutin peut dans de bonnes conditions garantir l'application de l'article 3 de la constitution. Par ailleurs, la multiplication des bureaux de vote et la variété de leur composition interdit à toute fraude systémique de rester discrète.

---

13   Historique des CSIRT (CERT en français) sur le site du CERT-FR https://www.cert.ssi.gouv.fr/csirt/

Ma conclusion finale est, pour que la loi et le règlement qui mettent en, place le vote électronique ne soient pas contraires à la constitution, il faut que la loi prévoit une annulation – y compris des années après – au simple motif que les garanties n'ont pas été apportées. Quoi qu'il en soit, contrairement au vote papier, le vote dématérialisé, y compris sur machines à voter, aux vues des connaissances actuelles, aujourd'hui ne peut apporter la garantie constitutionnelle du secret du vote.

J'en ai terminé et je vous remets cette contribution sous format écrite avec les références de ce que j'avance en vous remerciant de l'intégrer en annexe à vos publications et rapports. Je me tiens à votre disposition pour répondre à vos questions et je vous remercie de m'avoir écouté.

**COMPLEMENT : Sur les solutions « providentielles » et le risque de la « ré-identification »,**

Je veux mettre un point d'alerte particulier à propos des quelques personnes et quelques industriels qui assurent aujourd'hui proposer des solutions prétendant garantir la confidentialité du vote dématérialisé. Outre que cette garantie a été mise à bas mathématiquement comme je vous l'ai expliqué plus haut, je voudrais attirer votre attention sur une technologie à la mode et qualifiée de providentielle en matière de vote par internet.
Cette technologie, la blockchain (qui est celle qui tourne derrière la fameuse cryptomonnaie BitCoin) est présentée la plupart du temps comme la solution ultime pour le vote électronique. Comme pour les autres technologies généralement présentées en matière de vote électronique, elle ne règle aucunement les dangers précédemment évoqués, car elle opère sur une partie limitée de la chaîne de traitement et ne permet pas de traiter les risques associés aux terminaux des utilisateurs qui sont les appareils utilisés par les votants (ordinateurs et smartphones). Cette technologie, comme les autres, n'a pas non plus de moyen de convaincre tous les électeurs de la sincérité des opérations. Mais je voudrais surtout insister sur un danger majeur que cette technologie fait peser sur les votes : celui de la révélation du contenu intégral des votes le jour où une faille informatique est découverte dans un des protocoles de chiffrement utilisé dans cette blockchain. En effet, la blockchain est conçue pour être un registre non modifiable qui par construction doit rester public à long terme. Elle inscrit, et c'est dans son principe, dans un registre infalsifiable et ineffaçable, les traces chiffrées de chaque vote pour que ceux-ci ne puissent être altérés à posteriori.

Le risque de ré-identification est donc à prendre en compte en matière électorale. Il peut avoir des conséquences graves y compris des années après un scrutin si par exemple la nature des votes de tous les participants à une élection était révélée suite à l'exploitation d'une faille informatique, nouvellement découverte, qui concernerait un ancien enregistrement, obtenu de manière légale ou pas. C'est un risque majeur qui mérite d'être connu, et il n'y a aucune parade possible car l'avenir en matière de sécurité informatique est pavé de nombreuses surprises.

Pour ce genre de technologie, il y a un risque non négligeable que les votes puissent être un jour intégralement révélés.